\documentclass[10pt,english,twocolumn]{article}
\usepackage[T1]{fontenc}
\usepackage{geometry}
\usepackage{babel}

\usepackage{color}
\PassOptionsToPackage{normalem}{ulem}
\usepackage{ulem}

\usepackage{float}
\usepackage{graphicx}
\usepackage[numbers]{natbib}
\usepackage[unicode=true,pdfusetitle,
 bookmarks=true,bookmarksnumbered=false,bookmarksopen=false,
 breaklinks=true,pdfborder={0 0 1},backref=true,colorlinks=false]
 {hyperref}

\makeatletter
\usepackage{lmodern}
\date{}

\makeatother

\begin{document}

\title{ Uninformed sacrifice: evidence against long-range alarm transmission in foraging ants exposed to a localized perturbation}

\author{F. Tejera, A. Reyes  and E. Altshuler\thanks{Correspondence to: ealtshuler@fisica.uh.cu}}

\maketitle

\begin{center}
\emph{Physics Faculty, "Henri Poincar\`{e}" Group of Complex Systems, University of Havana, Cuba}
\end{center}

\abstract{It is well stablished that danger information can be transmitted by ants through relatively small distances, provoking either a state of alarm when they move away from potentially dangerous stimulus, or charge toward it aggressively. There is almost no knowledge if danger information can be transmitted along large distances. In this paper, we perturb leaf cutting ants of the species 
 {\it Atta insularis} while they forage in their natural evioronment at a certain point of the foraging line, so ants make a "U" turn to escape from the danger zone and go back to the nest. Our results strongly suggest that those ants do not transmit "danger information" to other nestmates marching towards the danger area. The individualistic behavior of the ants returning from the danger zone results in a depression of the foraging activity due to the systematic sacrifice of non-informed individuals.}


\section{Introduction}

One of the most amazing features of many species of ants is the
emergence of foraging lines that may span hundreds of meters from
the nest to the feeding sources \cite{Holldobler1990}. Such large
structures are particularly vulnerable, and the ability to balance
risk of death vs. value of food can provide a competitive advantage
to the colony. Unlike solitary animals, ants collect food not only
for their own consumption, but for the maintenance of the entire
colony as well \cite{Wilson2005,Boomsma2006}. In fact, the death of
a worker (of many workers) is not the end of reproduction and
therefore it has been viewed as a cost that the society is willing
to pay \cite{Bourke2008}. But how many ants the colony can afford
to sacrifice? How foraging ants react collectively to a source of
danger? These and other related questions have been rarely addressed
quantitatively in the literature \cite{Halley2001,Richardson2010a,Pinter-Wollman2013}.

A subject intimately linked to the matter is the transmission of danger
warming signals from informed to non-informed individuals. Danger
information is known to be transmitted at short distances --i.e, a 
few ant body lengths-- outside the nest by short-lived pheromone emissions,
body touching or even vibrations. It may result in either a state of
alarm when ants move away from potentially dangerous stimulus, or charge
toward it aggressively \cite{Wilson2005}. However, little is known about
the ability of ants to transmit danger information along large distances
--for example, from a certain point in the foraging line to the nest,
located a few meters a apart.

We have approached the problem by abducting leaf-cutter ants from
the species \emph{Atta insularis} at a given point of the foraging
trail in natural conditions, and quantifying ``long-range" effects
on several parameters of the foraging traffic in space and time. In
our experiments, approximately 50 percent of the ants are abducted
at a certain region of the foraging trail, and the rest are able to
avoid abduction: they make a ``U" turn (U-turn), and move back to
the nest. Our results suggest that those ants do not transmit danger
information to nestmates moving towards the abduction zone, so the
overall effect on the colony is that individuals are systematically
sacrificed in the attempt to maintain the foraging activity.

\section{Materials and Methods}
\label{sec:1}

Experiments were conducted on two colonies of the Cuban leaf-cutting
ant {\it Atta insularis}. The nests (that had not
been artificially modified in any way before our experiments) were
located under the pavement of one parking lot at the University of
Havana. Workers foraged every night on a garden located some 150
meters from the nest. The experiments were performed between the
22:00 and the 23:00 hours, corresponding to the peak of activity of
​​
activity in its steady state, in which the number of ants coming in
and out of the nest per unit time are equal and constant
\cite{Nicolis2013,Noda2006}. Fig. \ref{Experimental scheme} shows
our experimental setup: the nest's door is to the left, and the
foraging trail extends to the right. Two video cameras were used:
Camera 1 was near the door, immediately to its right. Camera 2 was 3
meters to the right of the door. During our experiments, two
interwoven lanes of ants were established: an {\it out-bound} one of
ants moving from the nest to the foraging area, and another of {\it
nest-bound} ants returning from the foraging area to the nest.

\begin{figure}[h!]
\begin{center}
\includegraphics[scale=0.45]{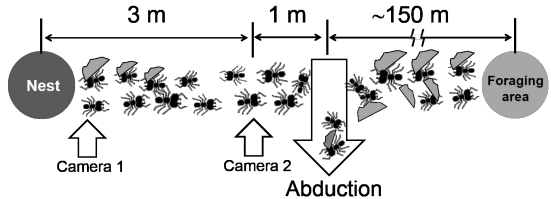}
\caption{{\bf Sketch of the experimental setup.} Camera 1 is near
the nest's door, Camera 2 is 3 m to the right of Camera 1, and the
abduction zone is 1 m to the right of Camera 2.\label{Experimental
scheme}}
\end{center}
\end{figure}

In a typical experiment, the unperturbed trail was filmed by the two
cameras for 25 minutes (this served as a baseline for stationary
activity). Then, we abducted ants using a vacuum cleaner in an area
of approximately 50~cm$^2$ located 1 m to the right of camera 2 (see
Fig. \ref{Experimental scheme}) either for 15, 25 or 30 minutes. The outbound ants coming
from the left that managed to escape from abduction simply returned
to the nest, while ants coming from the right could not cross the
abduction area to the left. After the abduction period, cameras 1
and 2 filmed the activity for another 20 minutes. Finally, ants were
returned to their natural environment, were they behaved normally.
Notice that perturbing by abduction has the advantage to ``isolate"
from the rest of the trail a section between the nest door and the
abduction area, which facilitates the quantitative analysis of the
data taken by Cameras 1 and 2, as we will see.

\section{Results}

Direct visual inspection and video analysis {\it a posteriori}
showed that most outbound ants reached the abduction area. There, 50
percent of the ants were abducted (i.e., sucked by the vacuum
cleaner). The other 50 percent stopped for a few seconds rising and
moving the antennae in the air, and then escaped the danger after
performing a U-turn to the nest. In spite of the fact that they
established numerous antennal contacts with the out-bound nestmates
moving in the direction of the abduction area, the latter did not
performed U-turns before reaching the abduction area.

\begin{figure*}[t!]
\begin{center}
\includegraphics[scale=0.78]{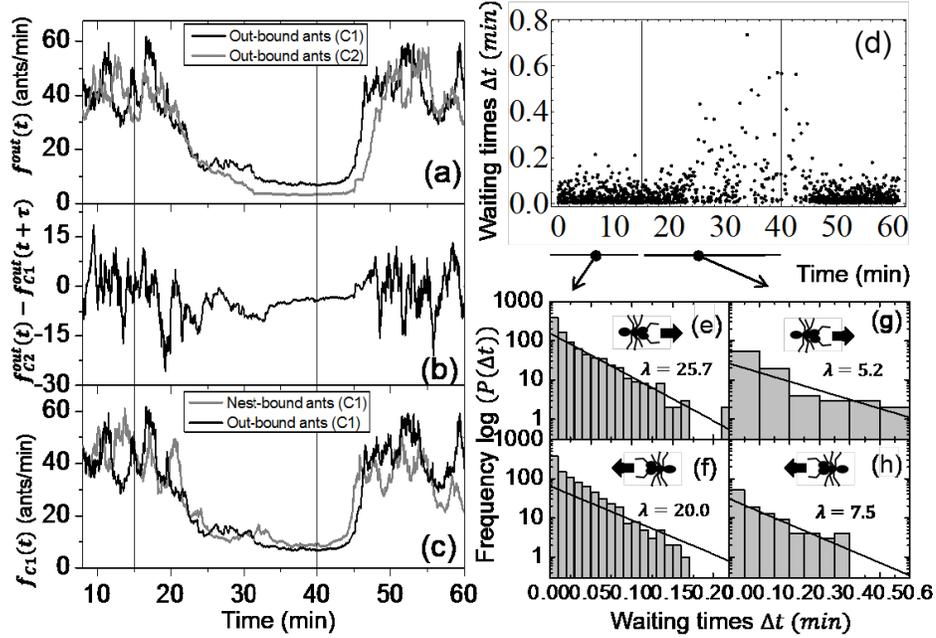}
\caption{{\bf Ants flows and waiting times distributions for an
abduction time of 25 min.} (a) Out-bound ants flow at Cameras 1 and
2. (b) Difference between the two curves shown in (a). (c) Out-bound
ants flow and Nest-bound ants at Camera 1. (d) Waiting times between
passage of consecutive out-bound ants, at Camera 1. (e) and (f)
Out-bound and Nest-bound waiting times distributions before
abduction, respectively. (g) and (h) Out-bound and Nest-bound
waiting times during the last 10 min of abduction, respectively. The
two vertical lines in (a) - (d) indicate the abduction period.
\label{fig: Flows and waiting times}}
\end{center}
\end{figure*}

That observation was corroborated quantitatively by measuring the
out-bound ant flows at cameras 1 and 2 and comparing them: Fig.
\ref{fig: Flows and waiting times} (a,b) illustrates the results for
one experiment. In order to estimate how many ants passed by camera
2 ($f_{C2}(t)$), relative to the total number of ants that passed by
camera 1 ($f_{C1}(t)$), we determined the differences between the
two flows. We smoothed out the ants flows by averaging every 300
points and take the difference as $f_{C2}(t)-f_{C1}(t-\tau)$. Here
$\tau$ represents {\it the delay between the flows} ($\tau=1.5$ min
for the experiment we are discussing here, which correspond to the
expected time an average ant moving at approximately 3 cm/s must
walk to cover a distance of 3 m). The fact that the difference
between the flows (Fig. \ref{fig: Flows and waiting times} (b))
during the abduction time is small and similar to the average
background before and after abduction corroborates quantitatively
the direct visual observation that no U-turns are {\it induced} on
the outbound ants by individuals returning from the abduction zone.
We also measured the Nest-bound and Out-bound ants passing by Camera
1.

Finally, we determined the temporal spacing between ants (or waiting
times \cite{Richardson2010a,Richardson2010,John09}). This parameter
is defined as the difference of passage time between an ant ({\it
i}) and its nearest nestmate ($i+1$): $\Delta t=t_{i+1}-t_{i}$. The
results are displayed in Figs. \ref{fig: Flows and waiting times}
(d-h), and suggest that the ``kinetics" of ants traffic far from the
abduction zone is not perturbed by the abduction process.

\section{Discussion}

As described before, when a typical out-bound ant reaches the
abduction zone, it may follow two fates: (a) being abducted or (b)
escaping after sensing danger. Choice (b) implies that the ants make
a U-turn at the abduction zone, and return to the nest. The
returning ants make many antennal contacts with their out-bound nest
mates. If the swarm acts collectively, it should be assumed that
danger information is spread through many foragers (perhaps through
chemicals \cite{Sasaki2014} or just by detecting a ``worrying"
decrease in the number of ants returning to the nest just by
counting antennal contacts). Those danger signals would induce
U-turns in the out-bound ants {\it before} reaching the abduction
zone, then increasing dramatically their survival probabilities. We
will call this the Cooperative hypothesis.

In Fig. \ref{fig: Flows and waiting times} we can see that the flow
of ants emerging from camera 1 is slightly bigger than the flow
passing by camera 2 (in normal conditions, some out-bound ants
perform U-turns independent of any perturbations in order to
``reinforce" the foraging pheromone track \cite{Evison2008}).
However, Fig. \ref{fig: Flows and waiting times} (b) shows that
U-turning remains constant and within the same range of values
throughout the whole experiment as compared to background
fluctuations, {\it including the abduction period} (averaged over
different experiments, $\Delta f =
<f_{C1}^{out}(t)-f_{C2}^{out}(t-\tau)>\mid_{no-abd}
-<f_{C1}^{out}(t)-f_{C2}^{out}(t-\tau)>\mid_{abd} = -3.48$ ants/min
$\pm$  4.65 ants/min, which is smaller than the fluctuations in
$\Delta f$: $\Delta f_{rms} = 5.97$ ants/min $\pm$ 1.44 ants/min,
where ``abd" means ``during the abduction period", ``no-abd" means
``out of the abduction period", and ``out" means ``outbound"). Since
the total number of out-bound ants between the two cameras is
conserved, we conclude that there is not an increment of U-turns
during the abduction period. Then, we reject the Cooperative
hypothesis: our ants act individualistically against danger, at
least on a large scale (i.e., at least more than 1 m away from
the abduction area).

Let us briefly de-tour from the subject of cooperation to find out
if ants keep a long memory of danger. A large number of ants
returning to the nest have directly experienced danger at the
abduction zone. One might expect that these individuals should stay
into the nest for a relatively long period, as an individualistic
protection mechanism. Our results suggest the opposite.  If we
analyze the flow of nest-bound and out-bound ants seen by camera 1
(Fig. \ref{fig: Flows and waiting times} (c)), we notice that the
graphs are very similar and the correlation coefficient between
 these two flow is $0.86 \pm 0.05$, for all the experiments. It indicates that there is not a
significant amount of ants that decide to remain inside the nest for
a longer time than usual because of the danger. This result suggests
that {\it foraging ants do not memorize danger information, at least
for a period of time longer than a few minutes.}

The distributions of waiting times for the out-bound and nest-bound
ants, can be described by a Poisson process, i.e. by exponential
distributions $P(t) = e^{-\lambda t}$. Figs. \ref{fig: Flows and
waiting times} (e-h) show the histograms for the waiting times
(including ants moving in both directions), fitted to exponential
distributions. Before the kidnapping the distributions for the
nest-bound and out-bound ants in both experiments are very similar,
indicating that the activity is stationary. During the abduction we
reduced the density of ants on the line, so we see longer waiting
times for the nest-bound and out-bound ants, implying a smaller
slope of the distribution (plotted in a log-linear graph). The
distribution of waiting times during the abduction for nest-bound
and out-bound ants, are also very similar (see Fig. \ref{fig: Flows
and waiting times} (g) and (h))  indicating no substantial changes
in the traffic between nest-bound and out-bound ants. This provides
extra evidence supporting our previous conclusions: {\it ants do not
keep a memory record of danger, and do not share danger information
with their nestmates.}

Finally, we underline some limitations of our experiments. (a)We
cannot check if ants escaping from a direct abduction attempt do
transmit danger information to the ones moving from the nest to the
abduction area, but {\it those ants do not use that information to
perform U-turns to avoid danger}: metaphorically speaking, we do not
know if ants just don't believe in ``conspiracy theory".(b) We have
only probed the ``long-range" consequences of possible transmission
of danger information: in principle, such information may be
transmitted locally (i.e., near the abduction area) {\it but} with
no effect on the overall foraging dynamics. (c) Similar experiments
during the {\it initial} foraging stage, or perturbing the nest near
its door could provoke a different collective response. This may
also happen using ``stronger" perturbations like a chemical
repellent \cite{Altshuler2005} of heating \cite{Boari2013}.

\section{Conclusions}

We have performed experiments where foraging ants are abducted at a
specific location of the foraging trail. Our results indicate that:

\begin {enumerate}

\item{Ants directly facing an abduction attempt that are able to
escape from it, perform a U-turn, and head back to the nest.}

\item{Ants escaping a direct abduction attempt act
individualistically: they do not transmit danger information along 
large distances, potentially useful to save the lives of ants moving
towards the danger zone by means of U-turns, or to recruit ants
from the nest in order to fight the external threat}

\item{Conclusions 1-3 suggest that, in the presence of a
spatially and temporally confined danger, there is not long-range
transmission of danger information, resulting in a "collective atempt"
to keep the foraging activity, even at the cost of many individuals}

\end {enumerate}

\section*{Acknowledgements}
We acknowledge support by K. Robbie and useful discussions with S. Nicolis and M. O. Magnasco. E. A. got inspiration from the late M. \'{A}lvarez-Ponte.
F. T. and A. R. acknowledge funding by the ``Henri Poincar\`{e}" Group of Complex Systems
E. A. acknowledges partial support through Total-ESPCI and Joliot-ESPCI ParisTech Chairs.
%
%
\bibliographystyle{}

\end{document}